\documentclass[prl,twocolumn,showpacs]{revtex4}

\usepackage{graphicx}
\usepackage{amsmath}

\usepackage{amsmath}
\usepackage{amsfonts}

\usepackage{tikz}

\def\be{\begin{equation}}
\def\ee{\end{equation}}
\def\ba{\begin{eqnarray}}
\def\ea{\end{eqnarray}}

\begin{document}

\title{Critical exponents of domain walls in the two-dimensional Potts model}

\author{J\'er\^ome Dubail$^{1,2}$, Jesper Lykke Jacobsen$^{2,3}$ and Hubert Saleur$^{1,4}$}
\affiliation{${}^1$Institut de Physique Th\'eorique, CEA Saclay,
91191 Gif Sur Yvette, France}
\affiliation{${}^2$LPTENS, \'Ecole Normale Sup\'erieure, 24 rue Lhomond, 75231 Paris, France}
\affiliation{${}^3$Universit\'e Pierre et Marie Curie, 4 place Jussieu, 75252 Paris, France}
\affiliation{${}^4$Department of Physics,
University of Southern California, Los Angeles, CA 90089-0484}

\date{\today}

\begin{abstract}

 We address the geometrical critical behavior of the two-dimensional
 $Q$-state Potts model in terms of the spin clusters (i.e., connected
 domains where the spin takes a constant value). These clusters are different
 from the usual Fortuin-Kasteleyn clusters,
 and are separated by domain walls that can cross and branch.
 We develop a
 transfer matrix technique enabling the formulation and numerical
 study of spin clusters even when $Q$ is not an integer. We further
 identify geometrically the crossing events which give rise to
 conformal correlation functions. This leads to an infinite series of
 fundamental critical exponents $h_{\ell_1-\ell_2,2\ell_1}$, valid for
 $0 \le Q \le 4$, that describe the insertion of $\ell_1$ thin and
 $\ell_2$ thick domain walls.

\end{abstract}



\maketitle


Many geometrical features of two-dimensional (2D) critical phenomena
are by now under complete control, thanks to the combined powers of
Conformal Field Theory (CFT) and Schramm-Loewner Evolution (SLE)
\cite{JJreview,BBreview}. This situation is epitomized by the
$Q$-state Potts model for which the Fortuin-Kasteleyn (FK) expansion
of the lattice model gives rise to a formulation in terms of clusters
and their surrounding loops (hulls). These loops behave like the SLE
trace in the continuum limit \cite{BBreview}, and viewing them as
contour lines of a (deformed) Gaussian free field leads to the Coulomb
Gas (CG) approach to CFT \cite{JJreview,Nienhuis}. Our understanding of the
critical properties of FK clusters and loops can be considered
almost complete.

And yet the properties of spin clusters themselves---i.e., the
connected domains with a constant value of the Potts spin---have as a
rule remained ill understood. This is particularly frustrating, since
those are the very clusters that one would observe in an actual
experiment on a magnetic alloy in the Potts universality class. The
case of the Ising model $Q=2$ is an exception to this rule
\cite{DuplantierSaleur,Vanderzande}, but this is due to its
``coincidental'' equivalence to the O($n$) vector model with
$n=1$. Indeed, defining the Ising spins on the triangular lattice, the
corresponding O($n$) model is described by self and mutually avoiding
loops on the hexagonal lattice, and such loops are readily treated by
CFT and SLE techniques.

\begin{figure}
 \includegraphics[width=0.23\textwidth]{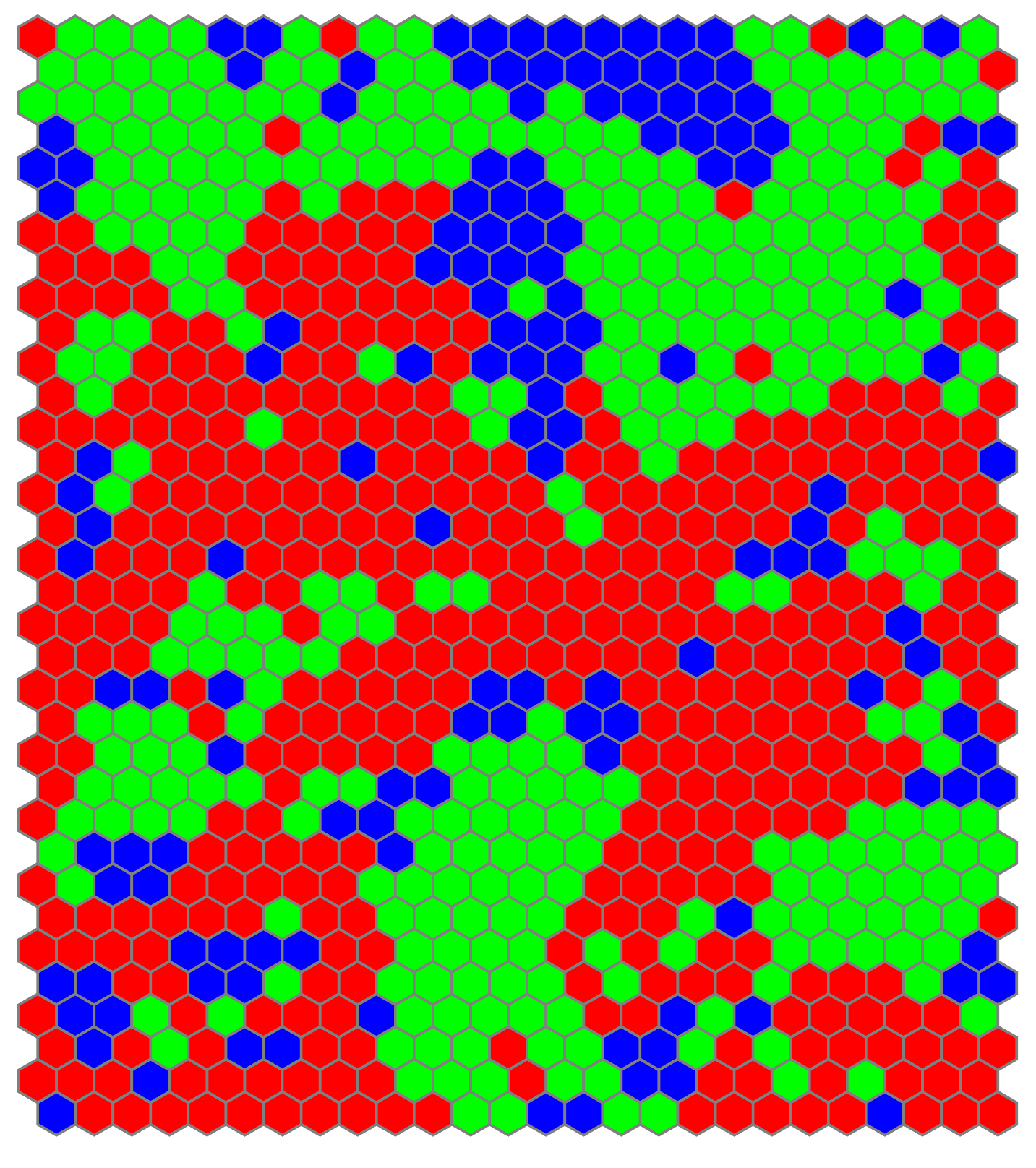}
 \includegraphics[width=0.23\textwidth]{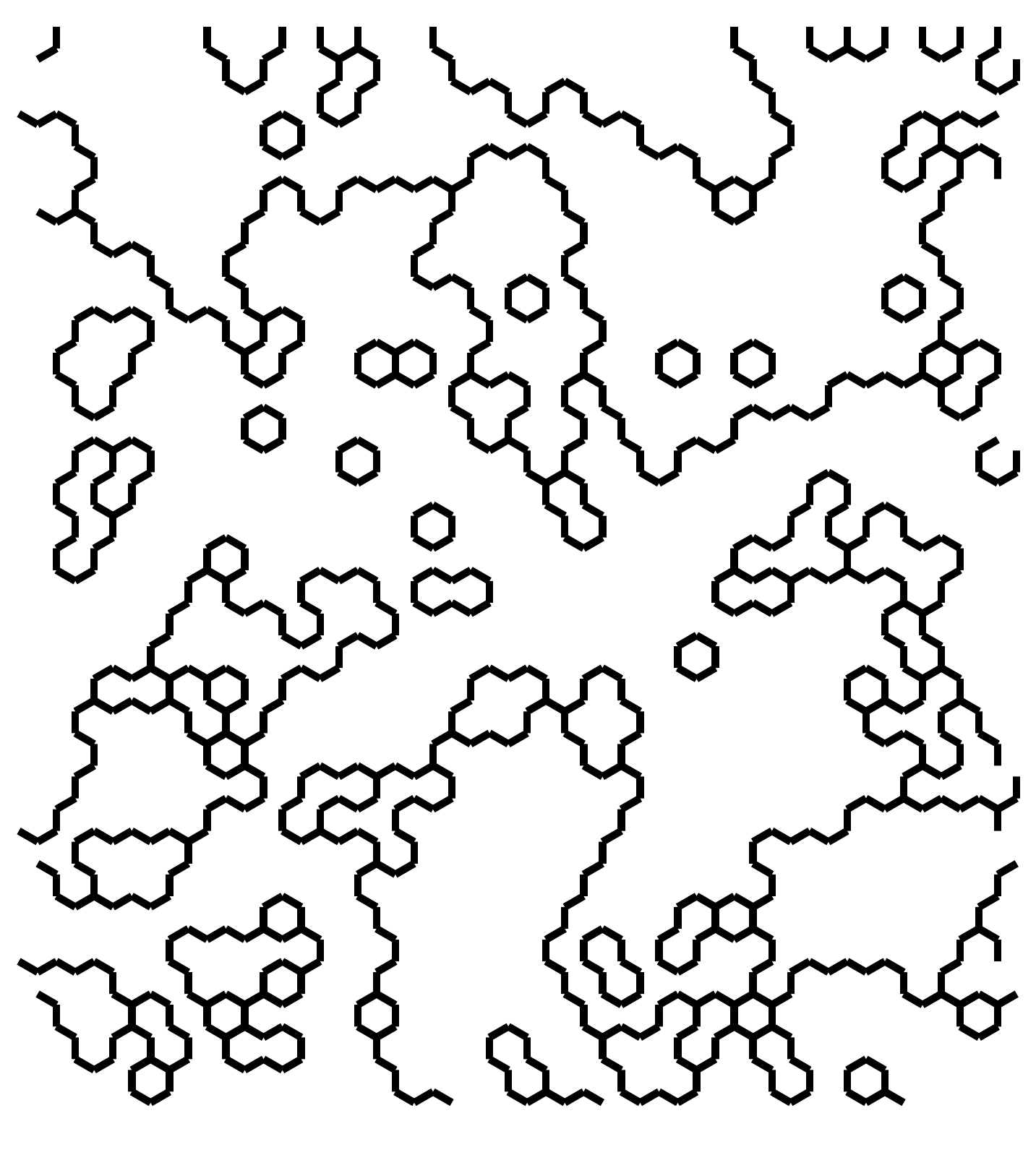}
 \caption{A configuration of the $Q=3$ Potts model, and the
   corresponding set of branching domain walls.}
 \label{fig:pottsDW}
\end{figure}

For general $Q$, the salient feature of Potts spin clusters is that
the domain walls separating different clusters undergo branchings and
crossings (see Fig.~\ref{fig:pottsDW}). These phenomena are however
absent for $Q=2$ with the above choice of lattice. It is precisely
these branchings and crossings that make the application of exact
techniques---such as CG mappings or direct Bethe ansatz
diagonalization---very difficult, if not impossible. The belief that
spin clusters are indeed conformally invariant for other values of $Q
\neq 2$ in the critical regime $0 \le Q \le 4$ has even been
challenged at times, but seems however well established by now
\cite{Coniglio,Qian}.

Some progress has been accomplished in the $Q=3$ case
\cite{Vanderzande, Janke} by speculating that the spin clusters in the
critical Potts model would be equivalent to FK clusters in the {\em
  tricritical} Potts model \cite{Stella,Janke}. This
equivalence has however not been proven, and is moreover restricted so
far to the simplest geometrical questions
\cite{Duplantierduality}. The equivalence can also be understood as a
relationship with the dilute O($n$) model \cite{CardyGamsa}.

The Potts model has been used recently to build a new class of 2D
quantum lattice models that exhibit topological order
\cite{FendleyModels}. Both FK clusters and domain walls
between spin clusters are important in these models.

Apart from issues of branching and crossing, another major hurdle in
the study of Potts spin clusters has come from the lack of a
formulation that can be conveniently extended to $Q$ a real
variable. In the case of FK clusters, this formulation led naturally
to the introduction of powerful algebraic tools via the Temperley-Lieb
(TL) algebra, and to the equivalence with the 6-vertex model---the
eventual key to the exact solution of the problem
\cite{JJreview}. Factors of $Q$ then appear naturally through a
parameter in the TL algebra, or---via a geometrical construction---as
complex vertex weights in the 6-vertex formulation. Also for spin
clusters can $Q$ be promoted to an arbitrary variable: the weight of a
set of spin clusters is simply the chromatic polynomial of the graph
dual to the domain walls.  From the point of view of the TL algebra,
the domain walls are composite (spin-1) objects, hence more
complicated than the FK loops.  Recent work on the related
Birman-Wenzl-Murakami (BWM) algebra \cite{ReadFendley,Fendley}
suggests that this formulation might be amenable to the standard
algebraic and Bethe ansatz techniques, although such a lofty goal has
not been achieved so far.

We report in this Letter major progress towards the understanding of
Potts spin clusters. Our results are of two kinds. On the one hand, we
develop a transfer matrix technique which allows the formulation and
numerical study of the spin clusters for all real $Q$. On the other
hand, we identify the geometrical events that give rise to conformal
correlations, and provide exact (albeit numerically determined)
expressions for an infinite family of critical exponents, similar to the
familiar ``$L$-legs'' exponents \cite{JJreview,Nienhuis} for FK
loops. Surprisingly, we find that geometrical properties of spin
clusters encompass all integer indices $(r,s)$ in the Kac table
$h_{r,s}$. An analytical derivation of our results appears for now
beyond reach, in part because the algebraic properties of our transfer
matrix are still ill understood. We do however provide some exact
results based on an approach which does not involve a CG mapping, but
rather the use of a massless scattering description.

\paragraph{Domain wall expansion.}

The $Q$-state Potts model is defined by the partition function
\be
 Z = \sum_{\sigma} \prod_{(ij) \in E}
 \exp \left( K \delta_{\sigma_i,\sigma_j} \right) \,,
 \label{Potts}
\ee
where $K$ is the coupling between spins $\sigma_i = 1,2,\ldots,Q$
along the edges $E$ of some lattice ${\cal L}$ (we use the square
lattice in the computations below, but the triangular lattice in the
figures). The Kronecker delta function $\delta_{\sigma_i,\sigma_j}$
equals 1 if $\sigma_i = \sigma_j$, and 0 otherwise.

\begin{figure}
 \includegraphics[width=0.23\textwidth]{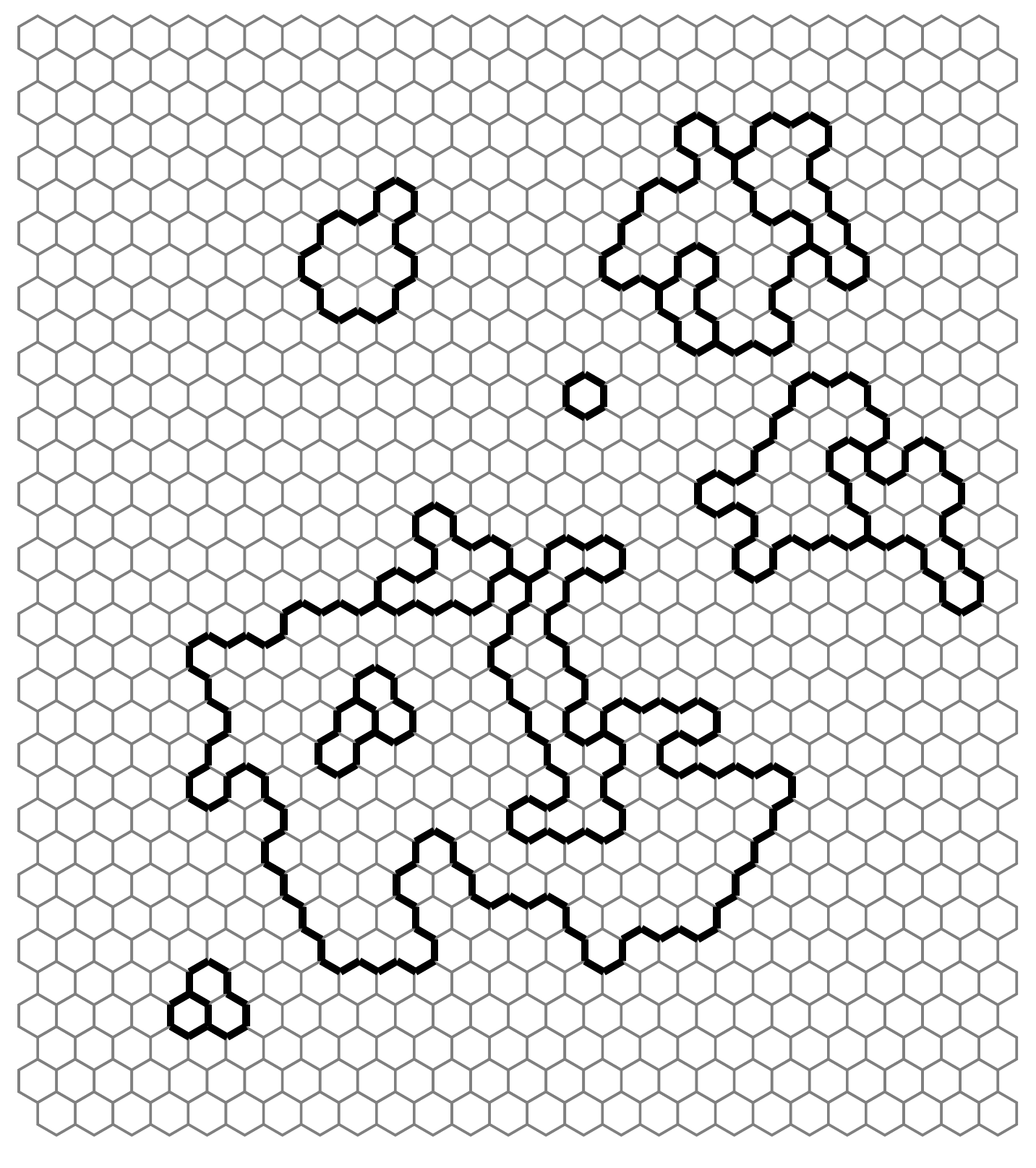}
 \includegraphics[width=0.23\textwidth]{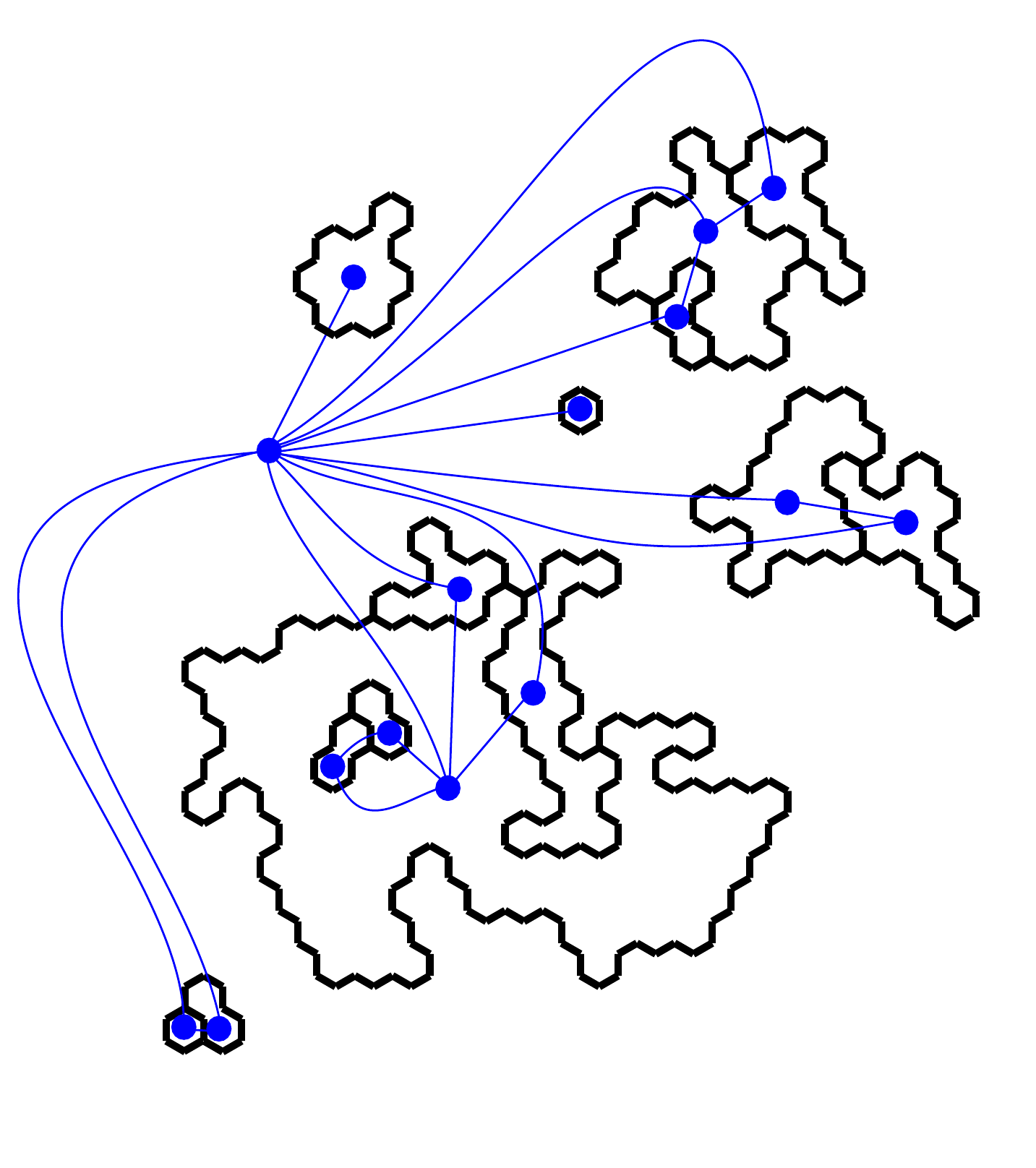}
 \caption{A domain-wall configuration corresponding to a graph $G$,
   and its dual graph $\hat{G}$.}
 \label{fig:DWgraph}
\end{figure}

The domain wall (DW) expansion of (\ref{Potts}) involves all possible
configurations of domain walls that can be drawn on the dual of ${\cal
  L}$ (see Fig.~\ref{fig:DWgraph}). A DW configuration is given by a
graph $G$ (not necessarily connected). The faces of $G$ are the spin
clusters.  Since we do not specify the color of each of these
clusters, a DW configuration has to be weighted by the chromatic
polynomial $\chi_{\hat{G}}(Q)$ of the dual graph $\hat{G}$.  Initially
$\chi_{\hat{G}}(Q)$ is defined as the number of colorings of the
vertices of the graph $\hat{G}$, using colors $\{1,2,\ldots,Q\}$, with
the constraint that neighboring vertices have different colors.  This
is indeed a polynomial in $Q$ for any $G$, and so can be evaluated for
any real $Q$ (but $\chi_{\hat{G}}(Q)$ is integer only when $Q$ is
integer).  For example, the chromatic polynomial of the graph
$\hat{G}$ on Fig.~\ref{fig:DWgraph} is $Q (Q-1)^7 (Q-2)^7$.
The partition function (\ref{Potts}) can thus be written as a sum over all
possible DW configurations 
\be
 Z = {\rm e}^{N K} \sum_{G} \, \left( {\rm e}^{-K} \right)^{{\rm length}(G)}
  \, \chi_{\hat{G}}(Q)
 \label{PottsDW}
\ee
where $N$ is the number of spins, and ${\rm length}(G)$ denotes the
total length of the domain walls.

\paragraph{The bulk DW exponents.}

The fundamental geometric object we consider is a connected part of a
domain wall that separates two clusters. One can ask how the
probability, that a certain number of such DW connect a small
neighborhood $A$ to another small
neighborhood $B$, decays when the distance $x$ between $A$ and $B$
increases. Each DW separates two spin clusters which connect $A$ and
$B$.
\begin{figure}[htbp]
 a. \,\includegraphics[width=0.22\textwidth]{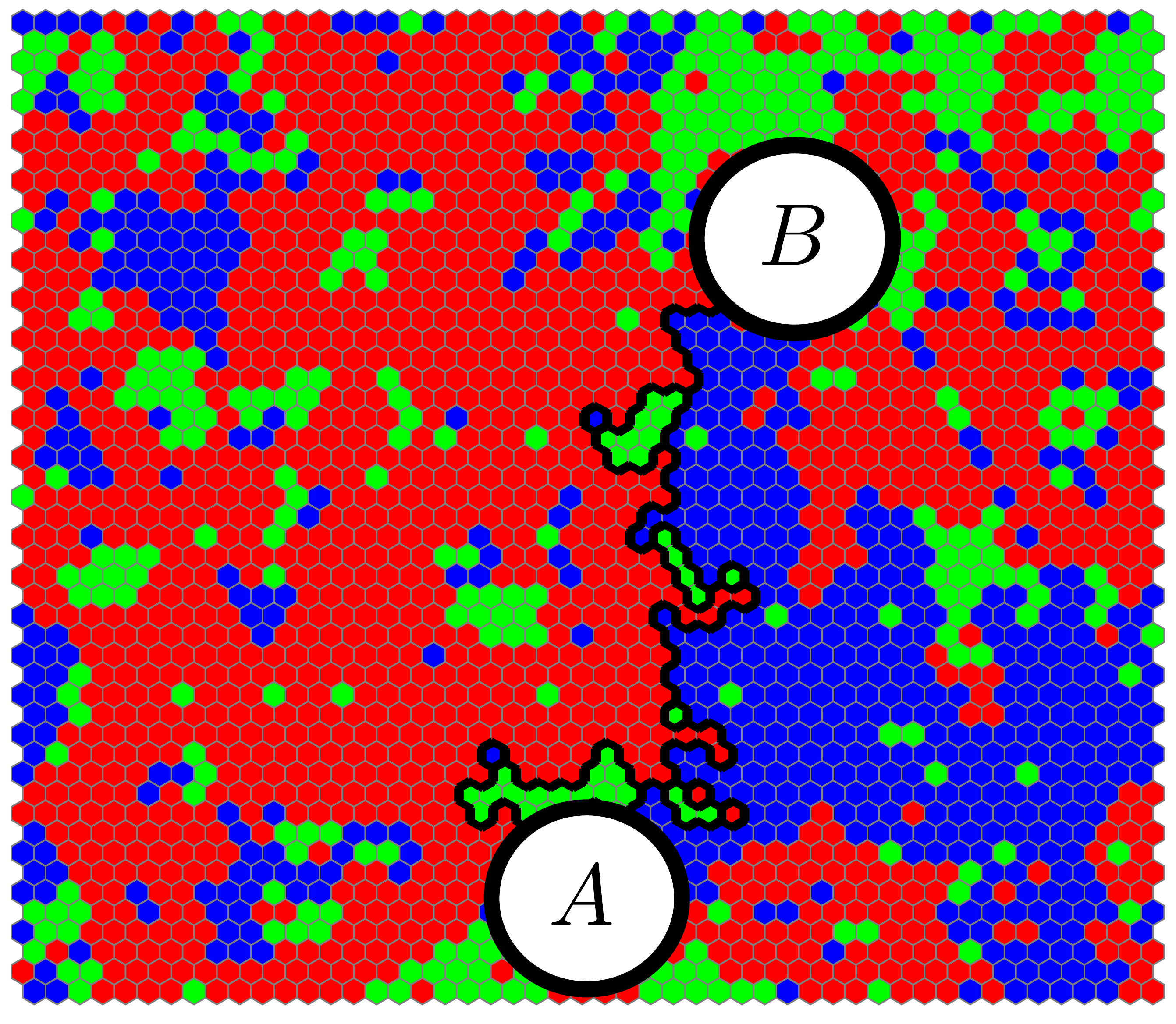}
    \includegraphics[width=0.22\textwidth]{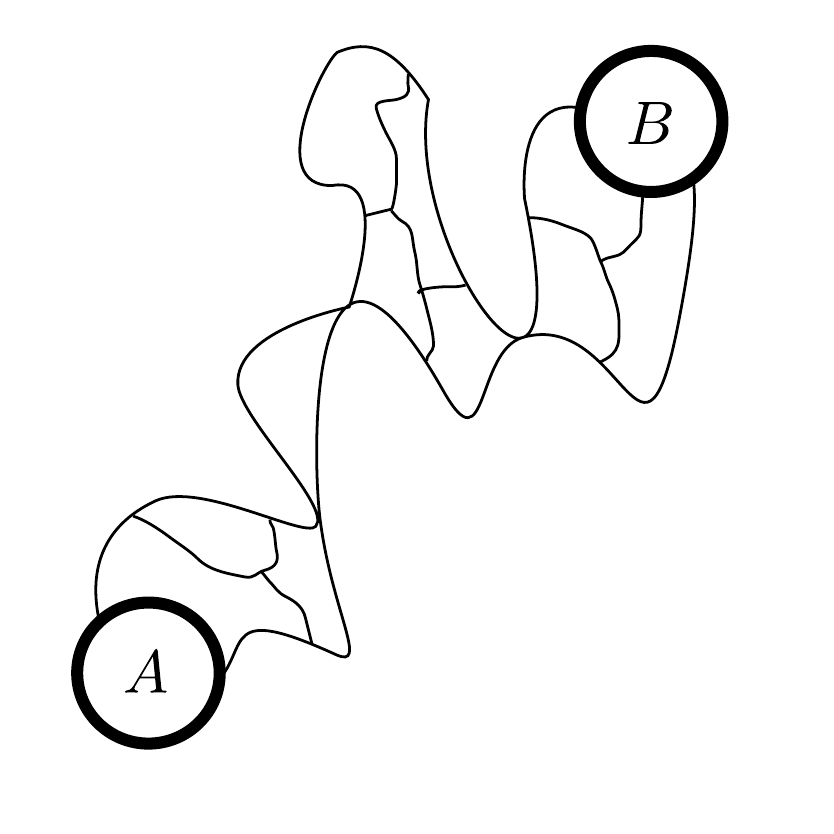}\\ \vspace{0.3cm}
 b. \, \includegraphics[width=0.22\textwidth]{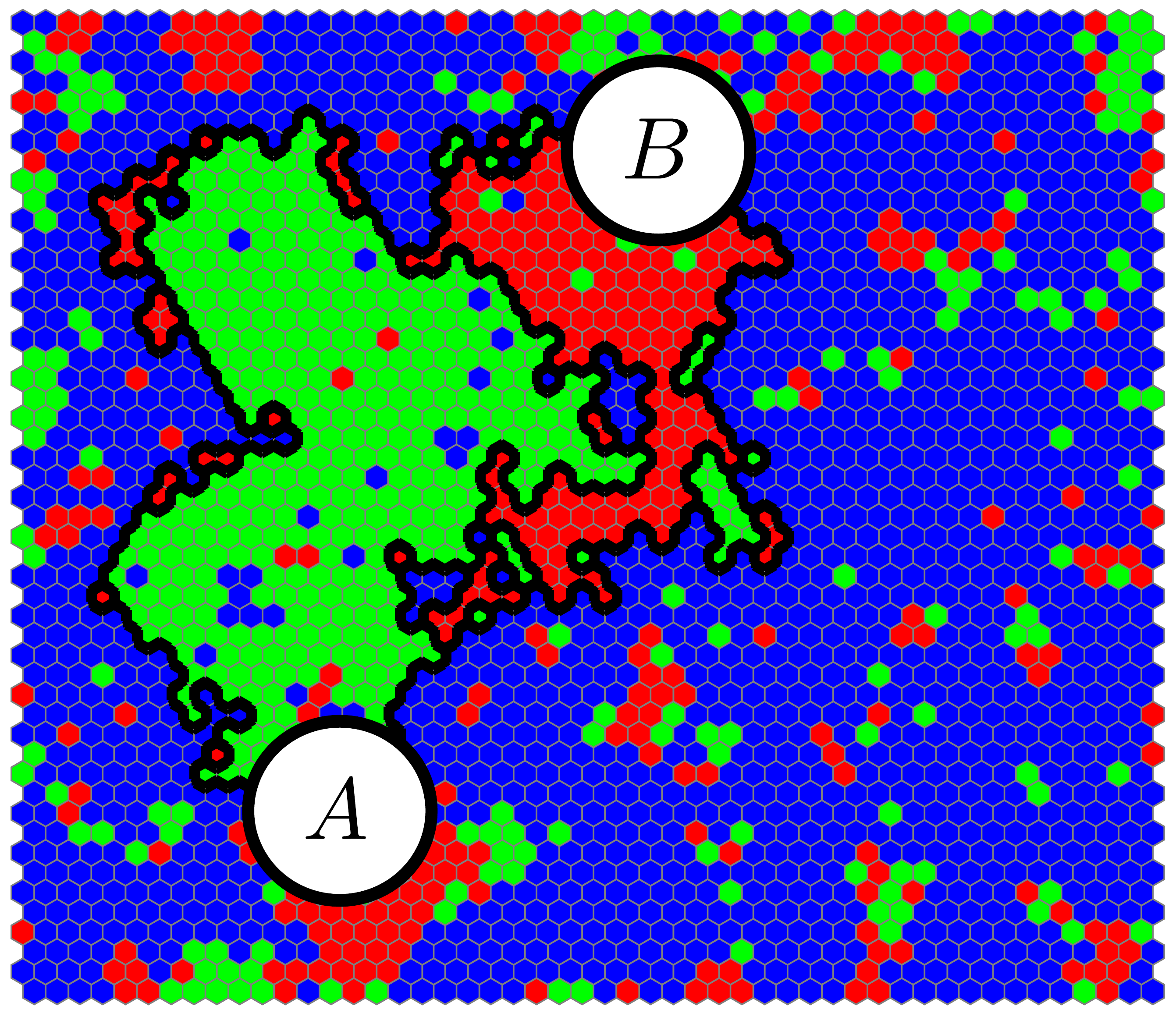}
    \includegraphics[width=0.22\textwidth]{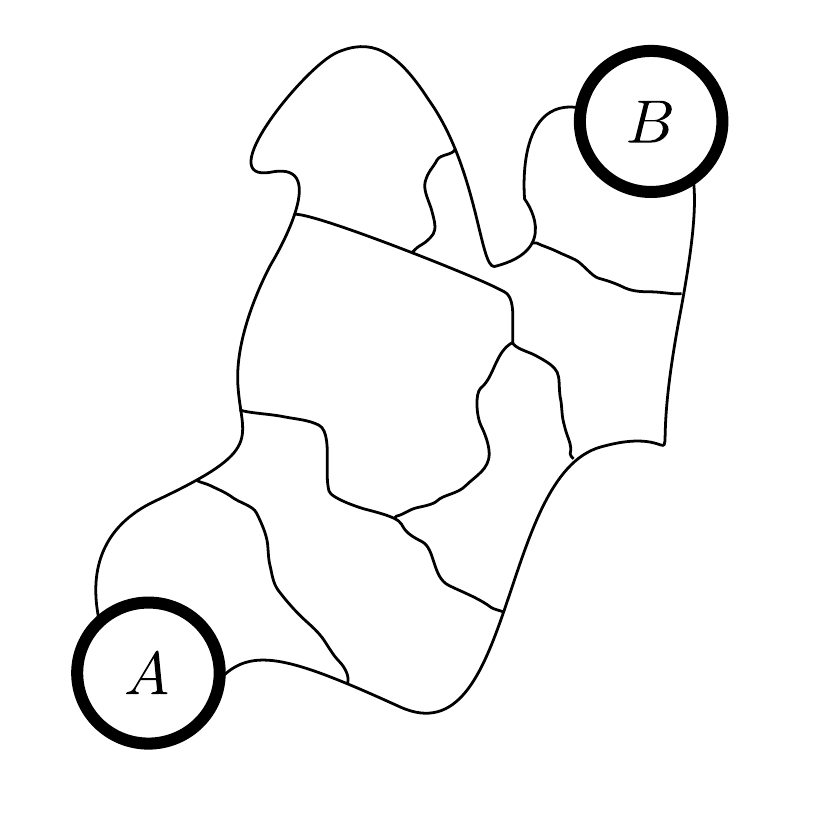}
 \caption{The two different types of DW: a \textit{thin} DW
   corresponds to the interface between two clusters of different
   colors (a), while for a \textit{thick} DW the two clusters have the
   same color. An illustration for the $Q=3$ Potts model is given
   (left) as well as a schematic picture for non-integer $Q$ (right).}
 \label{DW2types}
\end{figure}
There are in fact two types of such DW, depending on the relative
coloring of the two clusters that are separated. If the two clusters
have different colors, they can touch, so the DW is \textit{thin} (see
Fig.~\ref{DW2types}.a).  If the two clusters have the same color, then
they cannot touch (otherwise they would not be distinct), so the DW
has to be \textit{thick} (see Fig.~\ref{DW2types}.b).

We can now state the central claim of this Letter. Consider the 2D
Potts model for any real $Q$ in the critical regime $0 \le Q \le
4$. Then the probability $P$ that the two regions $A$ and $B$, with
separation $x \gg 1$, are connected by $\ell_1$ thin DW and $\ell_2$
thick DW decays algebraically in the plane with some exponent $h(Q,
\ell_1, \ell_2)$, viz., $P \propto x^{-4 h(Q,\ell_1
  ,\ell_2)}$. Equivalently, on a long cylinder of size $L \times \ell$
with $\ell \gg L$, and $A$ and $B$ identified with the opposite ends
of the cylinder, the decay is exponential: $P \propto {\rm e}^{- 4 \pi
  (\ell/L) \, h(Q, \ell_1, \ell_2)}$. Below we check this assertion
numerically, and we observe that the numerical values of the exponents
match the formula
\begin{equation}
 h(Q, \ell_1, \ell_2) = h_{\ell_1 - \ell_2, 2 \ell_1} \,,
 \label{mainconj}
\end{equation}
where we have used the Kac parametrization of CFT
\begin{equation}
 h_{r,s} = \frac{\left( r-s \, \kappa/4 \right)^2 - \left( 1 - \kappa/4 \right)^2}{\kappa}
 \label{Kac}
\end{equation}
and $2 \le \kappa \le 4$ parametrizes $Q = 4\left( \cos \frac{\kappa \,\pi}{4} \right)^2 \in [0,4]$.


\paragraph{Transfer matrix formulation.}

The DW expansion (\ref{PottsDW}) may appear unwieldy and difficult to
study in a Monte-Carlo simulation for non-integer $Q$.  However, it
can be tackled in a transfer matrix formalism that is no more
complicated than the one \cite{Blote82} routinely used in the study of
the FK clusters.

Consider a strip of the square lattice of width $L$ spins (boundary
conditions will be detailed later). The basis states on which the
transfer matrix $T$ acts contain one color label $c_i$ per spin.  By
definition, one has $c_i = c_j$ if and only if $\sigma_i = \sigma_j$
(i.e., the two spins on sites $i$ and $j$ have the same color). The
color labels $c_i$ contain less information than the spin colors
$\sigma_i$ themselves. For instance, any configuration in which the
first and third spins have the same color, no matter which one, and no
other spins have identical colors, is represented by
\begin{equation}
\label{eq:state}
\begin{tikzpicture}
        \filldraw (0,0) circle (0.7mm) node[above]{$c_1$};
        \filldraw (0.5,0) circle (0.7mm) node[above]{$c_2$};
        \filldraw (1,0) circle (0.7mm) node[above]{$c_1$};
        \filldraw (1.5,0) circle (0.7mm) node[above]{$c_4$};
        \draw (2,0) node{$\dots$};
        \filldraw (2.5,0) circle (0.7mm) node[above]{$c_L$};
\end{tikzpicture}
\end{equation}
For a row of $L=3$ vertices, and for any (non-integer) $Q$, there
are thus precisely five basis states
\begin{equation}
\label{basisL3}
 \begin{tikzpicture}
        \filldraw (0,0) circle (0.7mm) node[above]{$c_1$};
        \filldraw (0.4,0) circle (0.7mm) node[above]{$c_1$};
        \filldraw (0.8,0) circle (0.7mm) node[above]{$c_1$};
\end{tikzpicture}, \, \begin{tikzpicture}
        \filldraw (0,0) circle (0.7mm) node[above]{$c_1$};
        \filldraw (0.4,0) circle (0.7mm) node[above]{$c_2$};
        \filldraw (0.8,0) circle (0.7mm) node[above]{$c_1$};
\end{tikzpicture}, \, \begin{tikzpicture}
        \filldraw (0,0) circle (0.7mm) node[above]{$c_1$};
        \filldraw (0.4,0) circle (0.7mm) node[above]{$c_1$};
        \filldraw (0.8,0) circle (0.7mm) node[above]{$c_2$};
\end{tikzpicture} , \, \begin{tikzpicture}
        \filldraw (0,0) circle (0.7mm) node[above]{$c_1$};
        \filldraw (0.4,0) circle (0.7mm) node[above]{$c_2$};
        \filldraw (0.8,0) circle (0.7mm) node[above]{$c_2$};
\end{tikzpicture}, \, \begin{tikzpicture}
        \filldraw (0,0) circle (0.7mm) node[above]{$c_1$};
        \filldraw (0.4,0) circle (0.7mm) node[above]{$c_2$};
        \filldraw (0.8,0) circle (0.7mm) node[above]{$c_3$};
\end{tikzpicture}
\end{equation} 
Note that the last state would carry zero weight for $Q=2$, but
apart from that the number of basis states for any given $L$ will
be finite and independent of $Q$.

We can write $T$ as a product of elementary transfer matrices, each
represented symbolically as a rhombus surrounding a single lattice edge.
This edge links spins (shown as solid circles) on diametrically opposite
sites of the rhombus. On an $L=4$ square lattice with periodic boundary
conditions this reads
$T = \begin{tikzpicture}[scale=0.3]
        \draw (0,1) -- (0,0) -- (1,1) -- (2,0) -- (3,1) -- (4,0) -- (4,1) -- (3,0) -- (2,1) -- (1,0) -- (0,1) -- (0.5,1.5) -- (1,1) -- (1.5,1.5) -- (2,1) -- (2.5,1.5) -- (3,1) -- (3.5,1.5) -- (4,1);
        \filldraw (0,0) circle (1.4mm);
        \filldraw (1,0) circle (1.4mm);
        \filldraw (2,0) circle (1.4mm);
        \filldraw (3,0) circle (1.4mm);
        \filldraw (4,0) circle (1.4mm);
        \filldraw (0,1) circle (1.4mm);
        \filldraw (1,1) circle (1.4mm);
        \filldraw (2,1) circle (1.4mm);
        \filldraw (3,1) circle (1.4mm);
        \filldraw (4,1) circle (1.4mm);
\end{tikzpicture}$.

A rhombus
$\begin{tikzpicture}[scale=0.18] \draw (0,0) -- (1,1) -- (2,0) -- (1,-1) -- cycle;
        \filldraw (1,1) circle (2.4mm);
        \filldraw (1,-1) circle (2.4mm);
\end{tikzpicture}$
corresponds to a vertical edge, and acts on a basis state $s$ as
follows. If exactly $Q_s$ distinct color labels $\{c_k\}$ are used in
$s$, then the new color label $c_i'$ of the spin $\sigma_i$ can be
either unchanged $c_i'=c_i$ (with weight ${\rm e}^K$), or any one of the
other labels already in use $c_i'=c_k$ (each with weight $1$), or a
new one $c_i' \notin \{c_k\}$ (with weight $Q-Q_s$). Note that this
latter weight is in general non-integer, and is responsible for the
correct computation of the chromatic polynomial $\chi_{\hat{G}}(Q)$.
A rhombus
$\begin{tikzpicture}[scale=0.18] \draw (0,0) -- (1,1) -- (2,0) -- (1,-1)
 -- cycle;
	\filldraw (0,0) circle (2.4mm);
	\filldraw (2,0) circle (2.4mm);
\end{tikzpicture}$
adding a horizontal edge between vertices $i$ and $i+1$ corresponds
simply to a diagonal matrix, with a weight ${\rm e}^K$ if $c_i=c_{i+1}$, and
$1$ otherwise.

With these rules at hand, one can write the (periodic) $L=3$ transfer
matrix for arbitrary $Q$ in the basis (\ref{basisL3}) as an
instructive example:
$T \, =\, h_1 \cdot h_2 \cdot h_3 \cdot v_1 \cdot v_2 \cdot v_3$ with
\begin{equation*}
	v_1  =  \left(\begin{array}{ccccc} 
		{\rm e}^K & 0 & 0 & 1 & 0 \\
		0 & {\rm e}^K & 1 & 0 & 1 \\
		0 & 1 & {\rm e}^K & 0 & 1 \\
		Q-1 & 0 & 0 & {\rm e}^K +Q-2 & 0 \\
		0 & Q-2 & Q-2 & 0 & {\rm e}^K + Q-3 \\
 \end{array}  \right) 
\end{equation*}
%
and $h_1 = {\rm diag}({\rm e}^K,1, {\rm e}^K,1,1)$.
The remaining matrices can be obtained from those given by cyclic
permutations of the color labels. The reader can now check that
$T$ gives the same free energy as the FK transfer matrix, even when
$Q$ is non-integer.

However, to obtain the desired two-point correlation functions of DW,
the basis states (\ref{basisL3}) need to be endowed with some
additional information about the connectivity of the spin clusters. We
need to know whether two spins having the same label $c_i$ also belong
to the same cluster. Thus the states we use in the final transfer
matrix have the form
\begin{equation}
	\begin{tikzpicture}
	\draw[line width=2pt,gray] (1,0) arc (-180:0:0.25 and 0.5); 
	\filldraw (0,0) circle (0.7mm) node[above]{$c_1$};
	\filldraw (0.5,0) circle (0.7mm) node[above]{$c_2$};
	\filldraw (1,0) circle (0.7mm) node[above]{$c_1$};
	\filldraw (1.5,0) circle (0.7mm) node[above]{$c_1$};
	\filldraw (2,0) circle (0.7mm) node[above]{$c_5$};
\end{tikzpicture}  \qquad \quad
	\begin{tikzpicture}
	\draw[line width=2pt,gray] (0,0) arc (-180:0:0.75 and 0.5); 
	\draw[line width=2pt,gray] (1,0) -- (0.75,-0.5); 
	\filldraw (0,0) circle (0.7mm) node[above]{$c_1$};
	\filldraw (0.5,0) circle (0.7mm) node[above]{$c_2$};
	\filldraw (1,0) circle (0.7mm) node[above]{$c_1$};
	\filldraw (1.5,0) circle (0.7mm) node[above]{$c_1$};
	\filldraw (2,0) circle (0.7mm) node[above]{$c_5$};
\end{tikzpicture} 
\end{equation}
In the left state, the spins on vertices $3$ and $4$ are in the same
cluster, but not in the same cluster as the spin $1$. In the right
state the spins $1$, $3$ and $4$ are all in the same cluster. These two
states are different. In the transfer matrix evolution, each time two
neighbor vertices correspond to the same color, the corresponding clusters
are joined up.

The final transfer matrix thus keeps enough information, both about
the mutual coloring of the sites and about the connectivity of the
clusters, to give the correct Boltzmann weights to the different
configurations, even for non-integer $Q$, and to follow the evolution
of the boundary of a particular set of clusters. These boundaries are
precisely the domain walls (see Fig.~\ref{DW2types}).

\paragraph{Numerical results.}

We have numerically diagonalized the transfer matrix in the DW
representation for periodic strips of width up to $L=11$ spins. We
verified that the leading eigenvalue in the ground state sector
coincides with that of the FK transfer matrix, including for
non-integer $Q$.  As to the excitations, we explored systematically
all possible coloring combinations for up to $4$ marked spin
clusters, for a variety of values of the parameter $\kappa$. Finite-size
approximations of the critical exponents $h$ were extracted from the
leading eigenvalue in each sector, using standard CFT results
%
%
\cite{JJreview}. Final results (see Table~\ref{tab1}) for the
exponents were obtained by extrapolating those approximants to the $L
\to \infty$ limit. Note that (\ref{mainconj}) is only valid for
$(\ell_1,\ell_2)=(0,1)$ if the spin cluster is forbidden to wrap
around the periodic direction. Without that restriction we obtain
$h_{0,1/2}$.

\begin{table}
\begin{tabular}{r|rrrrr}
  $p=\frac{\kappa}{4-\kappa}$ &  $(2,0)$ &  $(0,2)$ &  $(3,0)$ &  $(2,1)$ &  $(0,3)$ \\ \hline
  $2$ &  2.01(1) &  5.99(2) &  2.97(4) &          &  8.94(2) \\ 
  $3$ &  4.01(1) &  7.99(2) &  6.02(2) &  8.04(2) & 11.98(3) \\
  $4$ &  5.93(2) & 10.01(2) &  8.89(5) & 10.93(5) & 15.05(4) \\
  $5$ &  7.77(4) & 12.09(8) & 11.6 (1) & 13.8 (1) & 18.2 (2) \\[0.1cm]
Exact & $2(p-1)$ & $2(p+1)$ & $3(p-1)$ & $3p-1$   & $3(p+1)$ \\
\end{tabular}
\caption{\label{tab1}
  Critical exponents corresponding to five different DW configurations
  $(\ell_1,\ell_2)$, as functions of the parameter $p = \frac{\kappa}{4-\kappa}$, along with
  the conjectured exact expression (\ref{mainconj}). The table entries
  correspond to $|\rho|$, when (\ref{Kac}) is rewritten as
  $h_{r,s} = (\rho^2-1)/(4p(p+1))$, with error bars in parentheses.} 
\end{table}

\paragraph{Analytical results.}

While it is far from obvious to derive (\ref{mainconj}) by CG methods,
minor progress can be achieved using a rather different set of
ideas. Indeed, we can learn about the dynamics of DW in the critical
theory by using known information about the low-temperature ($K>K_{\rm
  c}$) phase of the Potts model. Albeit non-integrable on the lattice,
the corresponding deformation by the operator $\Phi_{21}$ is
integrable in the continuum \cite{ChimZamo}. It can be described using
a basic set of kinks $K_{ab}$ separating two vacua, i.e., ordered
regions where the dominant value of the spin is $a$, resp.\ $b$. These
kinks scatter with a known S-matrix related to the BWM algebra
\cite{ReadFendley}.  Importantly, the dynamics conserves the number of
kinks: the process $K_{ab}K_{bc} \to K_{ac}$ is forbidden (as in any
elastic relativistic scattering theory), although kinks do appear as
bound states in kink-kink processes. Many properties of these kinks
can be calculated using integrability techniques. When the mass $m \to
0$ (i.e., $K\to K_c$), the S-matrix provides a ``massless scattering''
description \cite{FenSal} of some of the degrees of freedom of the
critical theory itself. It is not entirely clear what a kink, which is
well-defined for $K \gg K_{\rm c}$, becomes at $K_{\rm c}$, but it is
natural to expect that thick DW are described by the propagation of
two (or more) kinks such as $K_{ab}K_{ba}$. As for thin DW, the
potential existence of regions where they are reduced to a single
edge---that is a single kink---suggest they have to do with more
complicated processes involving two kinks merging into one. We will
not discuss them here.  It is an easy exercise to obtain the scaling
dimension of thick DW using the massless scattering
description. Indeed, the fact that the S-matrix satisfies relations
from the BWM algebra allows us to reexpress it in terms of the
$a_2^{(2)}$ or Bullough-Dodd S-matrix \cite{Efth}, for which the
thermodynamic Bethe ansatz was studied in \cite{SalWehe}. This
describes as well the dynamics of the field theory
\begin{equation*}
 S = \frac{1}{8\pi} \int {\rm d}^2x \,
 \left[(\partial_x\Phi)^2+(\partial_y\Phi)^2 +
 g(2 {\rm e}^{-\frac{\rm i}{\sqrt{2}} \beta\Phi} + 
 {\rm e}^{{\rm i}\sqrt{2}\beta\Phi})\right]
\end{equation*}
with $\beta^2 = \frac{\kappa}{4}$. Giving each kink the fugacity $Q-1$
produces the correct central charge $c=1-\frac{3}{2}
\frac{(\kappa-4)^2}{\kappa}$, where each kink has a $U(1)$ charge
equal to $0,\pm 1$. The scaling of the sector with charge $j$ produces
a gap $\Delta_j = \frac{j^2}{4 \kappa}$, so the leading dimension is
$\Delta_j-(1-c)/24 = h_{j/2,0}$.  This agrees with $h_{-\ell_2,0}$ with
$\ell_2=j/2$, so there are two kinks per thick DW.

\paragraph{Fractal dimensions.} The dimension of a spin cluster is
$d={\rm max} (2,\,2-2h_{0,1/2}) = {\rm max}\left(2,\, \frac{(8+\kappa)(8+3\kappa)}{32\kappa}\right)$, in agreement
with \cite{DuplantierSaleur,Vanderzande}. The boundary of a cluster has
dimension $d_{\rm b}=2-2h_{-1,0}=1+\frac{\kappa}{8}$. According to
(\ref{mainconj}), the set of points where a thin (resp.\ thick) DW has
minimal width (one resp. two lattice spacings) has dimension $d_1 = 2-2h_{2,4} =
\frac{3}{8 \kappa} (4-\kappa) (5 \kappa-4)$ resp.\ $d_2 = {\rm min}(0,\, 2-2h_{-2,0}) = 0$.
The fact that $d_1 \ge 0$ and
$d_2 = 0$ validates the epithets `thin' and `thick' for the scaling limit. For $Q=4$ (or
$\kappa=4$) we have $d_1=d_2=0$, so thin and thick DWs become
indistinguishable. Indeed one has $h_{\ell_1-\ell_2,2\ell_1} =
(\ell_1+\ell_2)^2/4$ in that case.  For the Ising model 
$Q=2$ (or $\kappa=3$) the absense of branchings means that
one thick DW equals two thin DW. Indeed $h_{\ell_1-\ell_2,2\ell_1} =
(4 \ell^2-1)/48$ with $\ell = \ell_1 + 2 \ell_2$ in that case, and
this agrees with the exponent $h_{\ell/2,0}$ for $\ell$ loop
strands \cite{JJreview, Nienhuis} in the dilute O($1$) model.


%

\smallskip

\paragraph{Acknowledgments.}

We thank G.\ Delfino and P.\ Fendley for discussions. This work was
supported by the Agence Nationale de la Recherche (grant
ANR-06-BLAN-0124-03).

\end{document}